\documentclass[useAMS,usenatbib]{mn2e}
\usepackage{amsmath,fleqn,graphicx,amssymb}
\usepackage{multirow}
\renewcommand{\[}{\begin{equation}}
\renewcommand{\]}{\end{equation}}
\def\p{\partial}\def\i{{\rm i}}
\def\ex#1{\left\langle#1\right\rangle}

\let\boldgrk=\gkvecten
\let\boldgrksc=\gkvecseven

\def\gkthing#1{{\mathchoice%
	{\hbox{{\boldgrk\char#1}}}
	{\hbox{{\boldgrk\char#1}}}
	{\hbox{{\boldgrksc\char#1}}}
	{\hbox{{\boldgrksc\char#1}}}}}

\def\vtheta{\gkthing{18}}

{\newif\ifnotend
\notendtrue
\def\veclist{ABCDEFGHIJKLMNOPQRSTUVWXYZabcdefghijklmnopqrstuvwxyz.}
\def\top#1#2.{#1}
\def\tail#1#2.{#2.}
\loop\expandafter\xdef\csname v\expandafter\top\veclist\endcsname%
{{\noexpand\bf\expandafter\top\veclist}}
\edef\veclist{\expandafter\tail\veclist}
\if\veclist.\notendfalse\fi\ifnotend\repeat}
\def\d{{\rm d}}
\def\cJ{{\cal J}}

\def\df{\textsc{df}}

\def\bolOm{\mbox{\boldmath$\Omega$}}
\def\vOmega{\bolOm}

\def\Gyr{\,\mathrm{Gyr}}
\def\Myr{\,\mathrm{Myr}}
\def\kpc{\,\mathrm{kpc}}

\def\kms{\,\mathrm{km\,s}^{-1}}

\def\pc{\,\mathrm{pc}}
\def\e{\mathrm{e}}
\def\Rc{R_\mathrm{c}}\def\RL{R_\mathrm{L}}
\def\fracj#1#2{{\textstyle{#1\over#2}}}

\def\tolJ{\textsc{tol}_J}

\def\vthetaT{\vtheta^{\rm T}}\def\vJT{\vJ^{\rm T}}

\def\TM{{\sc tm}}
\def\figref#1{Fig.~\ref{#1}}
\def\sos{{\sc sos}}
\def\tolJu{\kpc^2\Myr^{-1}}
\title[Resonant trapped orbits in our Galaxy]
{Managing resonant trapped orbits in our Galaxy}

\author[James Binney]{
  James Binney$^1$\thanks{E-mail: binney@thphys.ox.ac.uk}\\  
  $^1$Rudolf Peierls Centre for Theoretical Physics, 1 Keble Road,
  Oxford, OX1 3NP, UK
}

\begin{document}
\maketitle

\begin{abstract}
Galaxy modelling is greatly simplified by assuming the existence of a global
system of angle-action coordinates. Unfortunately, global angle-action
coordinates do not exist because some orbits become trapped by resonances,
especially where the radial and vertical frequencies coincide. We show that
in a realistic Galactic potential such trapping occurs only on thick-disc and
halo orbits (speed relative to the guiding centre $\ga80\kms$). We explain
how the Torus Mapper code (TM) behaves in regions of phase space in which
orbits are resonantly trapped, and we extend TM so trapped orbits can be
manipulated as easily as untrapped ones. The impact that the
resonance has on the structure of velocity space depends on the weights
assigned to trapped orbits. The impact is everywhere small if each trapped
orbit is assigned the phase space density equal to the time average along the
orbit of the DF for untrapped orbits. The impact could be significant with
a different assignment of weights to trapped orbits.
\end{abstract}


\section{Introduction} \label{sec:intro}

The volume and quality of the data we have to characterise both our own
Galaxy and many external galaxies has increased enormously over the last
decade, and continues to increase rapidly through technological advances such
as ESA's astrometric satellite Gaia and a new generation of integral-field
units such as ESO's instrument MUSE. Adequate exploitation of the superb data
now accumulating must involve the construction of intricate chemodynamical models
of galaxies that include fully dynamical dark matter and several populations
of stars of varying age and chemical composition.

A promising approach to the construction of such models involves specifying
the distribution functions (\df) of several populations $\alpha$ as analytic functions
$f_\alpha(\vJ)$ of action integrals, and then determining the gravitational
potential $\Phi(\vx)$ that these populations jointly generate. Once that has
been done, a prediction for essentially any observable quantity can be
extracted from the model, and the model can be fitted to one or more data
sets by adjusting the parameters in the \df s.

Over the last several years we have been pursuing this line of attack in the
context of modelling our Galaxy
\citep{JJB10,JJBPJM11:dyn,SaJJB13b,Piea14,SaJJB15:EDF,BinneyPiffl2015}. This
approach has been fruitful. For example, it revealed that the Local Standard
of Rest deduced from Hipparcos data and used for over a decade was
$\sim6\sigma$ in error \citep{JJB10}. It has also provided by far the
tightest constraints on the mass of dark matter interior to the solar radius
$R_0$ \citep{Piea14} and demonstrated for the first time that at
Galactocentric radii $R\la3\kpc$ baryons have materially modified the
phase-space density of dark matter \citep{BinneyPiffl2015}. \cite{SchoenPJM}
have used models in which a large number of chemically distinct stellar
populations each has a \df\ $f(\vJ)$ to explain the connection between
inside-out growth of our Galaxy and correlations between rotation velocity
and chemistry.

The work just described rests on the assumption
that realistic galactic potentials admit global angle-action variables.
Unfortunately, global angle-action variables only exist when a resonant condition
between the fundamental frequencies of an orbit never leads to the orbit
becoming trapped by the resonance. In typical galactic potentials some
orbits do become resonantly trapped, so these potentials do not admit global
angle-action coordinates. Given this inconvenient truth, could an approach to
galaxy modelling that predicates the existence of global angle-action coordinates
be misleading? Can angle-action variables be extended to cover resonantly
trapped orbits? The purpose of this paper is to address these questions.

A cornerstone of our work is provided by the Torus Mapper (\TM), a numerical
code that fits null tori to a given Hamiltonian
\citep{JJBPJM16}.\footnote{This code can be downloaded
from\hfil\penalty-10000
github.com/PaulMcMillan-Astro/Torus.}
 Using \TM\
one can construct a Hamiltonian $\overline{H}$ that closely approximates the given
Hamiltonian $H$ and admits global angle-action coordinates. The procedure is
as follows \citep{KaJJB94:PhysRev,JJBPJM16}. At each point $\vJ$ on a grid in
action space, \TM\ is used to construct a torus, that is functions
$\vx_\vJ(\vtheta)$ and $\vv_\vJ(\vtheta)$ that give the ordinary phase-space
coordinates in terms of the angle variables $\vtheta$. From these tori a
torus can be constructed for any point in action space by
interpolation. The Hamiltonian $\overline{H}(\vJ)$ is then defined to be the angle
average of the given Hamiltonian $H(\vtheta,\vJ)$ over this torus.  Clearly
the resulting tori are, by construction, the orbital tori of $\overline{H}$ and define
a global system of angle-action coordinates.  Once $\overline{H}$ and its angle-action
coordinates have been constructed, perturbation theory can be used to study
resonant trapping of orbits in $H$, but in the Galaxy modelling work cited
above we have simply ignored resonant trapping. In some sense ignoring
resonant trapping is equivalent to approximating our Galaxy's Hamiltonian $H$
by the integrable Hamiltonian $\overline{H}$.

In this paper we show how to proceed to the next level of approximation, in
which we recognise zones of missing actions in the basic action space, and
with each such zone associate a family of resonantly trapped orbits. 

In Section~\ref{sec:Phi} we introduce the model Galaxy potential that is used
throughout and recall the principles of surfaces of section. In
Section~\ref{sec:orbit} we explain why at any energy of interest  the
resonance $\Omega_r=\Omega_z$ will occur, and determine the critical value of
the peculiar velocity that a star with the angular momentum of the Sun must
have in order to become resonantly trapped. In Section~\ref{sec:resTM} we
show how \TM\ behaves when asked to produce a torus that has, in fact, become
trapped. In Section~\ref{sec:ptheory} we use Hamiltonian perturbation theory
to construct tori for trapped orbits. In Section~\ref{sec:MW} we use \TM\ to construct
different kinds of resonantly trapped orbits in a realistic Galactic
potential, and investigate the extent of resonant trapping in this potential.
We investigate the seriousness of the errors made by ignoring the existence of
resonant trapping, and explain how \TM\ can be used to determine the density
of stars in velocity space when resonant trapping is taking into account. 
Finally, in Section~\ref{sec:conclude} we sum up
and look to the future. An appendix explains in more detail why \TM\ makes a
sudden transition at the centre of a zone of missing actions.

\section{The Galactic potential}\label{sec:Phi}

We frame our discussion in the context of a gravitational potential that
\cite{PJM11:mass} fitted to a variety of data for our Galaxy. Specifically,
we adopt the ``best'' potential in that paper, which is generated by thin and
thick stellar discs, a flattened (axisymmetric) bulge and a spherically
symmetric dark halo. Its local circular speed is $v_c=239\kms$. Our
discussion would not differ materially, however, had we adopted any
reasonably realistic axisymmetric potential. To evaluate the potential and
its derivatives we use the {\sc falPot} code distributed in the \TM\ package,
which implements an algorithm described by \cite{WDJJB98:Mass}, and was
extracted from Walter Dehnen's {\sc falcON} package. 

\begin{figure}
\includegraphics[width=\hsize]{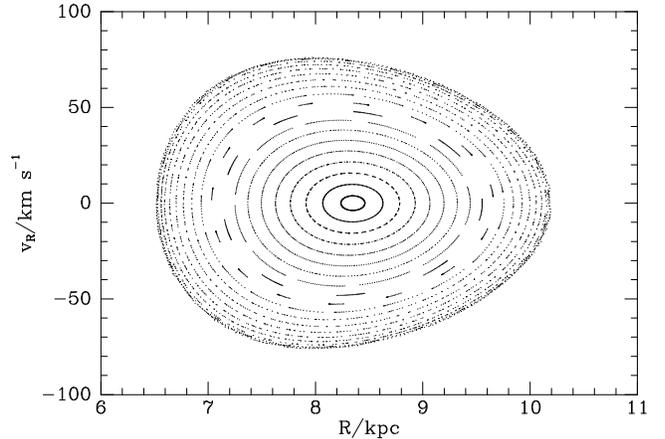}
\caption{Surface of section for orbits with the angular momentum $L_z$ of a
circular orbit at $R=8\kpc$ and the energy of this orbit plus
$\fracj12(0.32v_c)^2$.} \label{fig:two}
\end{figure}

Motion in an axisymmetric potential can be reduced to motion in the $(R,z)$
plane under the Hamiltonian
\[
H=\fracj12(p_r^2+p_z^2)+\Phi_{\rm eff}(R,z),
\]
 where
\[
\Phi_{\rm eff}(R,z)={L_z^2\over2R^2}+\Phi(R,z).
\]
 That is, we study motion with just two degrees of freedom under a
Hamiltonian that contains the angular momentum $L_z$ as a parameter. Deep
insight into such motion is provided by surfaces of section (\sos) in which
we plot a point in the $(R,p_R)$ plane every time the particle crosses the
equatorial plane moving upwards. All such ``consequents'' correspond to
phase-space point at a given energy $E$ at which $z=0$, so these points are subject
to two constraints on the four phase-space coordinates. If the orbit admits a
third integral, $I_3(R,p_R,z,p_z)$, this third constraint on the phase-space
coordinates restricts the consequents to one degree of freedom, and they lie
on a curve. \figref{fig:two} shows a surface of section for roughly the angular
momentum of the Sun. The points in this \sos\ clearly lie along a series of
curves, so an integral $I_3$ is respected by these orbits. Each curve
contains the consequents of a single orbit.

The curves in \figref{fig:two} form a bull's eye. At the centre would lie the
single point of the shell orbit $J_r=0$, which is the generalisation to a
flattened potential of a circular orbit that is inclined to the equatorial
plane. The curve of black points that runs around the edge, is formed by the
eccentric, planar orbit $J_z=0$. Hence, as one proceeds from the middle to
the edge, successive curves are generated by
orbits of increasing eccentricity and decreasing inclination.

\section{The 1:1 resonance near the Sun}\label{sec:orbit}

\begin{figure}
\includegraphics[width=\hsize]{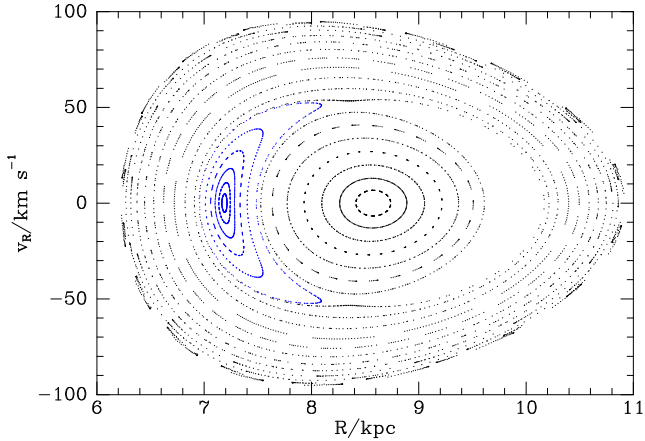}
\caption{Surface of section for the same angular momentum as \figref{fig:two}
but increased energy: we now add $\fracj12(0.4v_c)^2$ to the energy of the
circular orbit at $R=8\kpc$.}\label{fig:one}
\end{figure}

A resonance occurs whenever the fundamental frequencies
$(\Omega_r,\Omega_z,\Omega_\phi)$ of a quasi-periodic orbit satisfy a
relationship $\vN\cdot\vOmega=0$, where $\vN$ is a vector with integer
components.  The dynamical impact of a resonance rapidly decreases with the
modulus $|\vN|$, so by far the most important resonance for studies of our
Galaxy is the resonance $\vN=(1,-1,0)$ at which $\Omega_r=\Omega_z$. In our
adopted potential, the disc being massive and thin causes the radial epicycle
frequency $\kappa$ of circular orbits to be smaller than the vertical
epicycle frequency $\nu$ for $R<19\kpc$. Consequently, for disc orbits of low
inclination, $\Omega_z>\Omega_r$. By contrast, on the highly inclined orbits
of most halo stars $\Omega_z<\Omega_r$ because for a spherical system with a
flat circular-speed curve, $\Omega_r\simeq\surd2\Omega_\phi$ and
$\Omega_z=\Omega_\phi$. Hence given sufficient energy of motion in the
$(R,z)$ plane, somewhere along a sequence of orbits of decreasing $J_r$ and
increasing $J_z$, one will encounter the resonant condition
$\Omega_r=\Omega_z$.

To encounter the 1:1 resonance one has to consider more energetic orbits than
those plotted in \figref{fig:two}, which all have $\Omega_z>\Omega_r$. In the
present potential, orbits with the angular momentum of the Sun can become
resonantly trapped only if they have sufficient energy to pass the Sun at a
speed in excess of $82\kms$.

Whereas the orbits in the \sos\ of \figref{fig:two} were obtained by launching
particles from $(8,0)\kpc$ with a speed in the $Rz$ plane of $0.32v_c=76.8\kms$, the
orbits in \figref{fig:one} were obtained by launching particles with speed
$0.4v_c=96\kms$. Now there are orbits, with consequents plotted in blue, that are
trapped by the 1:1 resonance.  The innermost (black) curves are generated by
orbits with $\Omega_z<\Omega_r $, while the black curves that lie beyond the
blue curves are generated by orbits on which $\Omega_z>\Omega_r$.  Between
these regimes lie orbits of intermediate eccentricity on which
$|\Omega_r-\Omega_z|$ is so small that they become trapped. The orbits
generating the blue points always pass upwards through the plane at small
radii and downwards at large radii, so they circulate clockwise in the
$Rz$ plane. Other orbits, for which no consequents are plotted, circulate
in the opposite sense and would occupy the vacant region around
$R\simeq9.5\kpc$ -- we will discuss  these oppositely circulating orbits in
Section~\ref{sec:MW} below.

\section{TM at resonances}\label{sec:resTM}

\begin{figure}
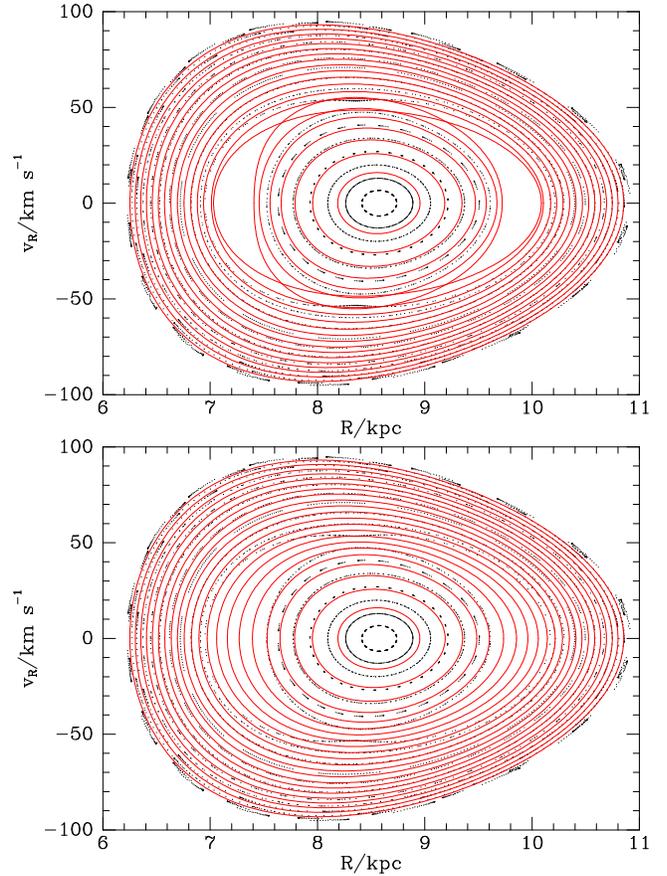

\includegraphics[width=\hsize]{plots/three.ps}
\includegraphics[width=\hsize]{plots/four.ps} \caption{Two examples of how
\TM\ generates tori in a region of orbit trapping. The black points are as in
\figref{fig:one}. the red points show the cross sections of tori generated by
\TM, using tolerance parameter $\tolJ=0.001\kpc^2\Myr^{-1}$ in the top panel
and $\tolJ=0.003\kpc^2\Myr^{-1}$ in the lower panel.}\label{fig:three}
\end{figure}

\figref{fig:three} shows two examples of how \TM\ generates tori at the
energy and angular momentum of \figref{fig:one}. Each panel is constructed by
finding $J_z$ for the shell orbit $J_r=0$ that has the correct energy, and
then incrementing $J_r$ in steps $\Delta$ while decrementing $J_z$ in steps
$(\Omega_r/\Omega_z)\Delta$ in order to keep the energy constant. For each
pair of actions a torus is found and used to draw a red curve. The only
difference in the procedure used to plot each panel of \figref{fig:three} is
that in the upper panel we specified tolerance $\tolJ=0.001\tolJu$, whereas for the
lower panel we specified $\tolJ=0.003\tolJu$.

The torus that wraps around the left side of the island of blue, trapped
orbits in \figref{fig:one} has $J_r=44.75\kpc\kms$, while the torus that
touches the right edge of the island has $J_r=28.13\kpc\kms$. Hence for the
studied energy, there are no orbits with $J_r$ in the range
$(28.13,44.74)\kpc\kms$; this is a zone of ``missing actions'' (ZoMA) in the
notation of \cite{BinneySII}.

The pattern of red curves in the upper panel of \figref{fig:three} is
unsatisfactory in that a few red curves cross other red curves. If each red
curve really were a cross section through a torus of the specified energy,
from a point of intersection one could complete the phase-space coordinates
$(R,v_R,z=0,v_z)$ by deducing $v_z>0$ from the energy, and on integrating the
orbit the consequents could lie on at most one of the red curves. Hence red
curves can intersect only if the Hamiltonian does not take the same,
constant, value on both the generating tori. Hence in \figref{fig:three} an
intersection of red curves implies that at least one of the responsible tori
is not an orbital torus.

When the actions specified are those of a trapped orbit, \TM\ cannot fit a
torus with the given actions exactly into the phase-space hypersurface $H=E$
because no such torus exists. Given a sufficiently large value of the
tolerance parameter $\tolJ$, it returns a torus that is smooth but fluctuates
in energy. Given a smaller value of $\tolJ$, it reduces
the fluctuations in energy just a little by returning a torus that depends
sensitively on the input actions.

We usually want the tori returned by \TM\ to be extensible to the basis of an
integrable Hamiltonian. Intersecting tori do not form such a basis, so they
should be avoided. The lower panel of \figref{fig:three} shows that in
this instance we can obtain tori that form a basis for the construction of an
integrable Hamiltonian simply by increasing the tolerance parameter to
$\tolJ=0.003\tolJu$.

\begin{figure}
\includegraphics[width=\hsize]{plots/five.ps}
\caption{As \figref{fig:three} but with a smaller value of the tolerance
parameter $\tolJ=.0005\tolJu$ and tori returned with ${\tt flag}\ne0$ rejected.}
\label{fig:four}
\end{figure}

A different strategy for avoiding intersecting tori is to use a small value
of $\tolJ$ but to reject tori returned by \TM\ with a non-zero value of the
variable {\tt flag}, indicating that the requested rms variation of the
Hamiltonian over the constructed torus was not achieved. \figref{fig:four}
illustrates this possibility. Now we have a consistent set of fits to all
un-trapped tori, and a clean gap left in the trapping region.

\subsection{Coefficients returned by TM}

\begin{figure}
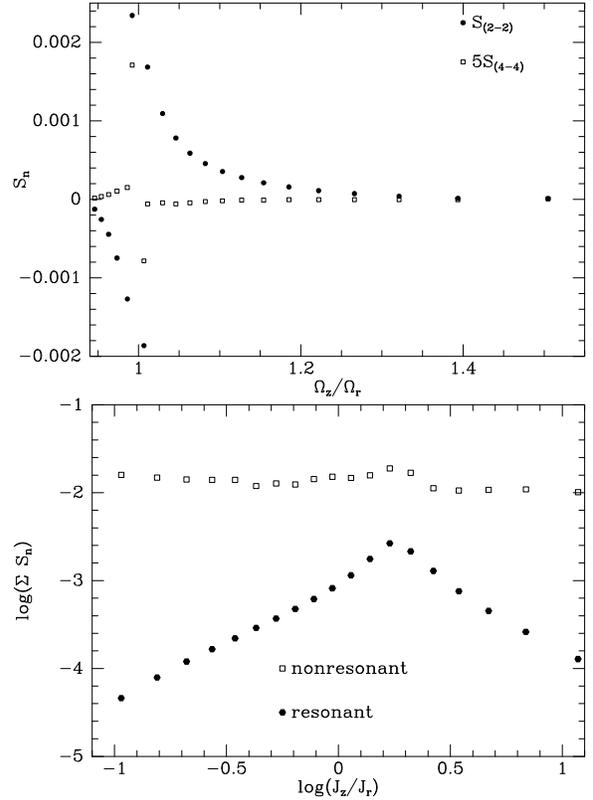

\begin{center}
\includegraphics[width=.9\hsize]{plots/Snplot.ps}
\includegraphics[width=.9\hsize]{plots/sumSn.ps}
\end{center}
\caption{Top: the values of $S_{(2-2)}$ (filled points) and $S_{4-4)}$ (open
points) on tori constructed by \TM\ for the surface of section
\figref{fig:four} plotted against frequency ratio. Below: sums of $|S_\vn|$
over all non-resonant  (open symbols) and resonant (filled symbols)
coefficients versus the ratio of actions.}
\label{fig:Sn}
\end{figure}

In this section we ask how \TM\ responds when asked to compute a torus that
has actually become trapped so it is impossible to drive the mean-square
value of the residual
Hamiltonian
\[
H_1(\vtheta,\vJ)\equiv H(\vtheta,\vJ)-\overline{H},
\]
 with $\overline{H}\equiv \ex{H}_\vtheta$, below some smallest, non-zero value.

\TM\ generates tori as images of the analytic torus of the isochrone
Hamiltonian under a canonical map. Normally the generating function of this map is
\citep{JJBPJM16}
\[\label{eq:defsS}
S(\vJ,\vthetaT)=\vJ\cdot\vthetaT+2\sum S_\vn(\vJ)\sin(\vn\cdot\vthetaT),
\]
where $\vthetaT$ denotes the angle variables of the isochrone Hamiltonian and
$\vn$ is a two-vector with integer components.  \TM\ minimises $\ex{H_1^2}$
with respect to the parameters of the isochrone Hamiltonian and the
coefficients $S_\vn$ in equation \eqref{eq:defsS} -- below we consider the
set of all the quantities to be adjusted to form the components of the vector
$\va$. The analytic Hamiltonian (in this case the isochrone Hamiltonian) is
referred to as the ``toy'' Hamiltonian, and the superscript T on $\vtheta$
in equation \eqref{eq:defsS} indicates that the components of $\vtheta$ are
the angle variables of the toy Hamiltonian.

When tori are computed for a progression of actions that includes resonantly
trapped tori, one expects the magnitudes of coefficients $S_\vn$ for which
$\vn\cdot\vOmega\to0$ to be large on either side of the resonance, reflecting
a growing need to distort the toy torus into the true torus. Moreover,
surfaces of section such as \figref{fig:four} suggest that these resonant
coefficients will have opposite signs on each side of the resonance, because
the toy tori are clearly stretched in opposite senses
(horizontally/vertically) on each side of the resonance. So we expect the
resonant $S_\vn$ to become large and positive as we approach the resonance
from one direction, to change sign in the ZoMA, and then to diminish from a
large negative value as we recede from the resonance. The upper panel of
\figref{fig:Sn} confirms this prediction by plotting against frequency ratio
$\Omega_z/\Omega_r$ the values of $S_{(2,-2)}$ and $S_{(4,-4)}$ along the
sequence of tori that make up the surface of section \figref{fig:four}. The
lower panel of \figref{fig:Sn}, which has a logarithmic axes, shows how strongly the importance
of the resonant coefficients grows as one approaches by plotting sums of
$|S_\vn|$ over resonant and non-resonant values of $\vn$.

\subsection{When do tori cross?}

\begin{figure}
\begin{center}
\includegraphics[width=.9\hsize]{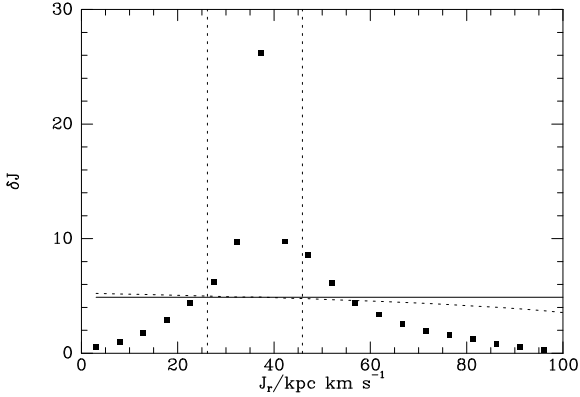}
\end{center}
\caption{The full and dashed lines show the increments $\delta J_r$ and
$\delta J_z$, respectively, between the tori shown in the upper panel of
\figref{fig:three}. The squares show twice the magnitude of the term
$2(2S_{(2,-2)}+4S_{(4,4)})$ in equation \eqref{eq:JtoJ} that changes
sign as one passes through the resonance. The vertical lines mark the
edges of the ZoMA.}\label{fig:cross}
\end{figure}

If we want to use tori returned by \TM\ to define an integrable Hamiltonian,
it is vital that neighbouring tori do not cross in any surface of section.
From the  generating function \eqref{eq:defsS} we have that
\[\label{eq:JtoJ}
\vJT=\vJ+2\sum_\vn \vn S_\vn(\vJ)\cos(\vn\cdot\vthetaT).
\]
Two tori, $\vJ$ and $\vJ+\delta\vJ$ will cross if a single phase-space point
$(\vthetaT,\vJT)$ can be reached from both tori. The tori most likely to
cross are those on either side of the resonance $\vN\cdot\vOmega=0$ because,
as we have seen, then $S_\vN(\vJ)\simeq-S_\vN(\vJ+\delta\vJ)$. Since
$\cos(\vN\cdot\vthetaT)$ is unity somewhere on the tori and $S_\vn$ often
changes sign across a resonance, crossing is likely
if
\[
|\delta\vJ|<4\biggl|\sum_\vN\vN S_\vN\biggr|,
\]
 where the sum is over resonant $\vN$. In \figref{fig:cross} the nearly
horizontal lines show the differences in  $J_r$ and $J_z$ between adjacent
tori in the upper \sos\ of \figref{fig:three} and the dots show twice the
value of the  sum
\[\label{eq:ressum}
2(2S_{(2,-2)}+4S_{(4,-4)}),
\]
 which changes sign across the resonance. The region in which the dots lie
above the horizontal lines, implying the likelihood of tori crossing,
coincides quite well with the ZoMA, the boundaries of which are marked by the
vertical lines. Evidently, if the tori are required to form an integrable
Hamiltonian, the sum \eqref{eq:ressum} should be prevented from exceeding
$\sim\frac12|\delta\vJ|$.

\section{Perturbation theory}\label{sec:ptheory}

When on some torus $\vJ_0$ the resonance condition $\vN\cdot\vOmega(\vJ_0)=0$
is satisfied, one should use perturbation theory to investigate the
possibility of resonant trapping. In the vicinity of the resonance, the angle
variable
\[\label{eq:thetas}
\theta_1'\equiv\vN\cdot\vtheta
\]
 will evolve slowly and we make a canonical transformation to new
angle-action variables $(\vtheta',\vJ')$ that include this variable. A
suitable generating function is
\[
S'(\vtheta,\vJ')=J_1'\vN\cdot\vtheta+J_2'\theta_2+J_3'\theta_3.
\]
 Indeed, then from $\vtheta'=\p S/\p\vJ'$ one recovers equation \eqref{eq:thetas} and
$\theta'_{2,3}=\theta_{2,3}$. From $\vJ=\p S/\p\vtheta$ we find
\begin{align}\label{eq:defsJp}
J_1&=N_1J'_1     &                 J'_1&=J_1/N_1\cr
J_2&=N_2J'_1+J'_2&\leftrightarrow\qquad  J'_2&=J_2-J_1{N_2/N_1}\cr
J_3&=N_3J'_1+J'_3&                 J'_3&=J_3-J_1{N_3/N_1}.
\end{align}
It is worth noting that 
\[\label{eq:Omzero}
{\p H\over\p J'_1}={\p H\over\p\vJ}\cdot{\p\vJ\over\p J'_1}
=\vOmega\cdot\vN=0,
\]
so when we vary $J'_1$ while holding $J'_{2,3}$ fixed, we are considering
orbits of a common energy, as in a surface of section.

We Fourier expand the Hamiltonian in the new angle variables
\[\label{eq:expH}
H(\vtheta',\vJ')=\overline{H}(\vJ')+\sum_{\vn\ne0}h_\vn\e^{\i\vn\cdot\vtheta'},
\]
 where $h_\vn\ll \overline{H}$ because the $\vJ'$ are close to the actions of $H$.
The new actions have the equations of motion
\[
\dot\vJ'=-{\p H\over\p\vtheta'}=-\i\sum_\vn \vn
h_\vn\e^{\i\vn\cdot\vtheta'}.
\]
 We average these equations over $\theta'_2$ and $\theta'_3$ on the grounds
that these variables move through a complete cycle in times that are much
shorter than the smallest time $|\vJ|/h_\vn$. Since the equations of motion
for $J'_{2,3}$ only contain terms with non-trivial dependence on
$\theta'_{2,3}$, the right sides of these equations vanish after averaging,
and we conclude that $J'_{2,3}$ are effective constants of motion: on short
timescales they wiggle slightly, but on long timescales they do not change.
From equations \eqref{eq:defsJp} we see that the constancy of $J'_{2,3}$
implies that when $N_{2,3}\ne0$, any variation in $J_1$ will be reflected in
a complementary variation in $J_{2,3}$.

The averaged equations of motion can be derived from the Hamiltonian obtained
by averaging equation \eqref{eq:expH} over $\theta'_{2,3}$:
\[\label{eq:Hbar}
H(\theta'_1,\vJ')=\overline{H}(J_1')+\sum_{n\ne0}h_n(J'_1)\e^{\i n\theta'_1},
\]
 where $h_n\equiv h_{(n,0,0)}$ and we have omitted references to the constant
actions $J'_{2,3}$. Since this a time-independent Hamiltonian, the motion
occurs on the curve in the $(\theta'_1,J'_1)$ plane on which $H=I_1$, a constant. A
good approximation to this motion can be obtained by Taylor expanding the
functions of $J'_1$ in equation \eqref{eq:Hbar}. However, before we do so we
exploit the
reality of $H$ to write
\[\label{eq:H1d}
H(\theta'_1,\vJ')\simeq\overline{H}(J_1')+2\sum_nh_n(J'_1)\cos(n\theta'_1+\psi_n),
\]
 where the $\psi_n$ are the phases of the $h_n$. We expand $\overline{H}$ and $h_n$ to second
order in 
\[
\Delta\equiv J'_1-J'_{01},
\]
 where $J'_{01}$ is the primed action
of the resonant torus. Since a constant term in $H$ can be discarded and we
know that $\p\overline{H}/\p J'_1=0$ on the resonant torus
(eq.~\ref{eq:Omzero}), we replace
$\overline{H}$ by $\fracj12 G\Delta^2$, where
\[
G\equiv{\p^2\overline{H}\over\p J'_1{}^2}={\p\overline{\Omega}'_1\over\p
J'_1}.
\]
The Taylor series for $h_n(J'_1)$,
\[
h_n(J'_1)=h_n^{(0)}+h_n^{(1)}\Delta+\fracj12 h_n^{(2)}\Delta^2+\cdots
\]
cannot be simplified in this way, so the
equation $H=I_1$ becomes
\begin{align}\label{eq:Hquad}
0&=\Bigl(\fracj12G+\sum_nh_n^{(2)}\cos(n\theta'_1+\psi_n)\Bigr)\Delta^2\cr
&+2\Bigl(\sum_nh_n^{(1)}\cos(n\theta'_1+\psi_n)\Bigr)\Delta\cr
&+\Bigl(2\sum_nh_n^{(0)}\cos(n\theta'_1+\psi_n)-I_1\Bigr).
\end{align}
In this approximation we can determine $J'_1(\theta'_1)$ simply by solving
the quadratic equation \eqref{eq:Hquad} for $\Delta$ given $\theta'_1$. Given the
fluctuating value of $J'_1$ and the constants $J'_{2,3}$, we can recover
the complete action-space coordinates from  equations \eqref{eq:defsJp}. In
general all three components of $\vJ$ oscillate but in such a way that $H$ is
to leading order constant.

If we retain only one value of $n$ and neglect $h_n^{(1)}$ and $h_n^{(2)}$,
equation \eqref{eq:Hquad} reduces to the energy equation of a pendulum, and
this is traditionally used to discuss resonant trapping
\citep[e.g.][]{Chirikov1979}. \cite{Kaasalainen_res} demonstrated the merit
retaining $h_n^{(1)}$ and $h_n^{(2)}$. In the application to the Galaxy good
results can be obtained with just the dominant value of $n$, but perceptibly
better results are obtained at trifling extra cost by retaining two values of
$n$, so that is what we do.

\subsection{Action and angle of libration}

The range through which $\theta'_1$ oscillates is set by the condition that
the quadratic for $\Delta$ has real roots:
\begin{align}\label{eq:discr}
\Bigl[\sum_nh_n^{(1)}\cos(n\theta'_1+\psi_n)\Bigr]^2&\ge
\Bigl(\fracj12G+\sum_nh_n^{(2)}\cos(n\theta'_1+\psi_n)\Bigr)\cr
&\times\Bigl(2\sum_nh_n^{(0)}\cos(n\theta'_1+\psi_n)-I_1\Bigr).
\end{align}
In our application to the Galaxy below we have $G<0$. This being so, and
bearing in mind that the $h_n$ are by definition all non-negative,
$\theta'_1$ oscillates around $\theta_1'=-\psi_n/n$ for the dominant value of
$n$. The largest permitted value of $I_1$ is set by adopting
the equals sign in equation \eqref{eq:discr} with
$\cos(n\theta'_1+\psi_n)=1$  for the dominant value of $n$, and similarly for
the smallest permitted value of $I_1$.

Rather counter-intuitively,  $I_{1\,\rm min}$, the smallest value of $I_1$,
corresponds to the orbit that has the largest-amplitude librations around the
resonant orbit. The reason $I_1$ does not function like a conventional energy
is that $G$, which in equation \eqref{eq:Hquad} plays the role of mass, is
negative.

Each value of $I_1$ corresponds to an action $\cJ$ that quantifies the
extent to which a trapped orbit oscillates around the trapping torus. $\cJ$
is straightforwardly computed as
\[
\cJ={1\over2\pi}\oint\d\theta'_1 J'_1(\theta'_1),
\]
 where the dependence of $J'_1$ on $\theta'_1$ is obtained from equation
\eqref{eq:Hquad}. Since $\theta'_1$ increases from its minimum to its maximum
value with $\Delta$ given by the larger root of the quadratic
\eqref{eq:Hquad} and returns to its minimum value with $\Delta$ given by the
smaller root, $\cJ$ is give by the difference of the roots $\Delta$
integrated over the range of $\theta'_1$.

On a trapped orbit,  $\theta'_1$ is not an angle variable, although
$\theta'_{2,3}$ are angle variables. Since  the missing angle variable evolves linearly in
time, it is
\[
\vartheta(\theta'_1)=2\pi{\int_0^{\theta'_1}\d\theta'_1/\dot\theta'_1\over
\oint\d\theta'_1/\dot\theta'_1},
\]
 where from Hamilton's equation and equation  \eqref{eq:H1d} we have
\[
\dot\theta'_1=G\Delta+2
\sum_n\left(h_n^{(1)}+h_n^{(2)}\Delta\right)\cos(n\theta'_1+\psi_n).
\]
 In this equation $\Delta$ is by the quadratic equation \eqref{eq:Hquad} a function
of $\theta'_1$.

\begin{table*}
\caption{Public methods of {\tt resTorus}}\label{tab:resTor}
\begin{center}
\begin{tabular}{ll}
{\tt resTorus(Tgr,Jrgr,n,Jb,G,hn)}&
{\hsize=.5\hsize\raggedright\vtop{{\tt Tgr} and {\tt Jrgr} pointers to array
of {\tt n} untrapped tori and their radial actions. {\tt Jb} actions of underlying resonant torus,
{\tt G} derivative of $\Omega'$, {\tt hn} pointer to $h_{2,-2}$, $h_{4,-4}$
and their
derivatives on torus {\tt Jb}.}}\\
{\tt I,Imin,Imax}&The value of $I_1$ on this torus and the smallest and
largest permitted values of $I_1$.\\
{\tt setI(I)}&Sets value of $I_1$\\
{\tt librationAction()}&Returns action of libration.\\
{\tt librationOmega()}&Returns frequency of libration.\\
{\tt FullMap(A)}&Returns $(R,z,\phi,v_R,v_z,v_\phi)$ at given angles.\\
{\tt SOS(ofile,n)}&Places in {\tt ofile} {\tt n} consequents in $(R,v_R)$ SoS.\\ 
{\tt SOSr(ofile,n)}&As {\tt SOS} but for the orbit rotating counterclockwise
in the $Rz$ plane.\\
{\tt containsPoint\_Ang(Rzphi,A,t1)}&
{\hsize=.5\hsize\raggedright\vtop{Returns number of times torus passes through
point {\tt Rzphi} with {\tt A} and {\tt t1} pointers to arrays where
corresponding angles
and values of $\theta_r-\theta_z$ are deposited.}}\\
{\tt containsPoint(Rzphi,V,A,t1)}&As above but including velocities of
visits.
\end{tabular}
\end{center}
\end{table*}

\subsection{Addition of a class to TM}

A class {\tt resTorus} has been added to \TM\ to create and manipulate
trapped tori. The public elements of this class are listed in
Table~\ref{tab:resTor}. The key methods {\tt FullMap} and {\tt containsPoint}
are very similar to the corresponding methods of the class {\tt Torus}
described by \cite{JJBPJM16}.  The class currently specialises to the case of
trapping by the $1:1$ resonance in the $Rz$ plane. By default is produces
tori for libration about $\psi=0$ or $\pi/2$ according to the sign of $h_2$.
The torus for libration about $\psi=\pi$ can be recovered by the
transformations $\theta_r\to 2\pi-\theta_r$, $\theta_z\to\pi-\theta_z$, which
imply $v_R\to-v_R$, $v_z\to-v_z$ and $\theta'_1=\pi-\theta'_1$. Tori for
libration around $\psi=-\pi/2$ can be generated by the mapping
$\theta_z\to\theta_z+\pi$, which implies $z\to-z$, $p_z\to-p_z$.

\section{Application to the Galaxy}\label{sec:MW}

We now apply the above apparatus to the resonance generated by the Galactic
disc, for
which $\vN=(1,-1,0)$, so $J_1'=J_r$ in this case.  We start by using the tori
fitted by \TM\ with $\tolJ=0.003\tolJu$ like those shown in the lower panel
of \figref{fig:three} to define the unperturbed Hamiltonian
$\overline{H}(\vJ)$.  Fourier analysis of the true Hamiltonian on the
resonant torus shows that all non-vanishing coefficients $h_n$ have even
$n_z$ and are real. Hence in the equations above, we should
consider terms with $n=2,4,6\ldots$. The largest of these terms is $h_2$ and
the next largest term is $h_4$. Since $h_4$ is not always negligible compared
to $h_2$, we include both $h_2$ and $h_4$.  

For values of $L_z$ corresponding to circular orbits at $\Rc\la4.5\kpc$,
$\psi_2=\pm\pi$, whereas when $\Rc(L_z)\ga4.5\kpc$, $\psi_2=0$ or $2\pi$.
Hence for low values of $L_z$ orbits librate around $\theta_1'=\pm\pi/2$, and
for higher values of $L_z$ orbits librate around $\theta_1'=0$ or $\pi$.

Since $\theta_r=0$ corresponds to pericentre and $\theta_z=0$ as the star
passes through the plane, $\psi_2=\pm\pi$ implies that the $R$ and $z$
oscillations are either in phase or in antiphase, so in the $Rz$ plane the
star moves from bottom left to top right, or top left to bottom right
depending on the adopted sign of $\psi_2$.

When $\psi_2=0$ the trapping closed orbit passes up through the plane at
pericentre and descends at apocentre, ensuring that the star has a well
defined sense of circulation in the $(R,z)$ plane. When $\psi_2=2\pi$, the
sense of rotation is reversed.

From the value taken by $h_2$ on the fitted resonant torus and an estimate of
$G$ obtained by fitting a line to the values of $\Omega'_1$ on the fitted
tori, we estimate the half width of the trapped region as
\[
\Delta_0=2\sqrt{2h_2\over|\d\overline{\Omega}'_1/\d J'_1|},
\]
 which is the value of $\Delta$ implied by equation \eqref{eq:Hquad} for
$I_1=I_{1\,\rm min}$ when only $n=2$ is retained and $h_2^{(1)}$ and $h_2^{(2)}$
are neglected. \TM\ with $\tolJ=0.003\tolJu$ is then used to fit the non-resonant
tori $J'_1=J'_{01}\pm\Delta_0$ that bound the trapping region.  Twenty tori
that run through the trapping region are obtained from these bounding tori by
linear interpolation. On these tori we Fourier analyse the Hamiltonian, and
fit natural cubic splines to the Fourier amplitudes $\overline{H}$, $h_2$ and
$h_4$ as functions of $J_r$. From these splines we obtain
$\Omega_1'=\d\overline{H}/\d J_r$ and re-determine the value of $J_r$ on the
resonant torus. The value of $J_r$ on the
resonant torus and the spline fits are now used to obtain the values $G$,
$h_n^{(0)}$, $h_n^{(1)}$, and $h_n^{(2)}$ required by the perturbation
theory. From this theory we update our estimates of the boundaries of the
trapping region, use \TM\ to fit new bounding tori, and by Fourier analysis
we recompute the numbers required by perturbation theory.

\begin{figure}
\begin{center}
\includegraphics[width=.9\hsize]{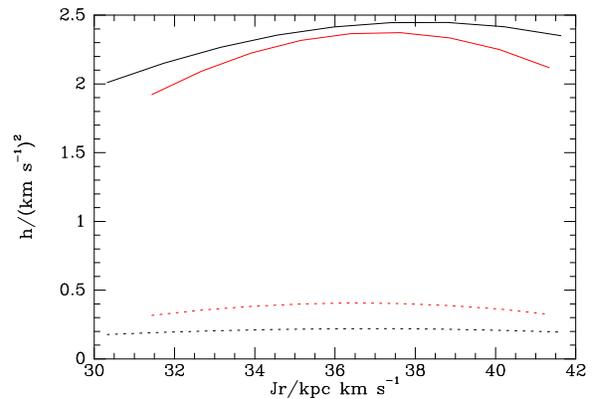}
\end{center}
\caption{Full curves: $h_2$ as a function of $J_1'=J_r$ when interpolating
between tori fitted either side of the trapping region with different values
of the precision parameter $\tolJ$. The red curve is for
$\tolJ=0.001\kpc^2\Myr^{-1}$ and the black curve is for
$\tolJ=0.003\kpc^2\Myr^{-1}$. Dashed curves: the corresponding plots of
$h_4(J_1')$.}\label{fig:h2h4}
\end{figure}

\figref{fig:h2h4} shows the values of $h_2$ and $h_4$ on tori obtained by
interpolating between tori fitted either side of the trapping region with
different values of the error parameter $\tolJ$. The red curves are for the
smaller value of $\tolJ$, so $h_4$ is increased by
using a smaller error parameter when fitting the tori. In fact the values of
$h_4$ shown in \figref{fig:h2h4} are sufficiently large that it becomes
just worthwhile to include in the perturbing Hamiltonian the terms
$2h_4\cos(4\theta_1')$. The biggest impact of including them is on orbits
that lie close to the parenting closed orbit.

\begin{figure}
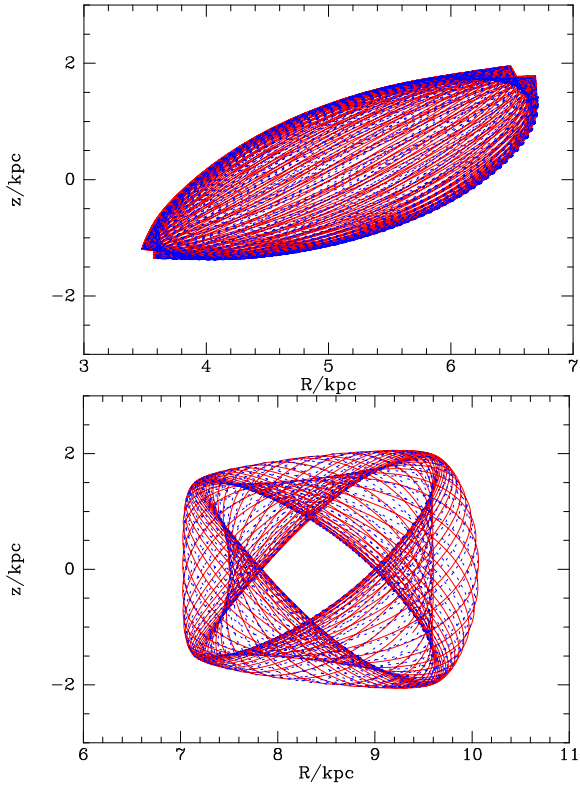

\begin{center}
\includegraphics[width=.9\hsize]{plots/full_int_orb0.ps}
\includegraphics[width=.9\hsize]{plots/full_int_orb.ps}
\end{center}
\caption{In red two resonantly trapped orbits reconstructed from perturbation
theory. In blue the orbits obtained by integrating the full equations of
motion from a point on the reconstructed orbit. The upper orbit has a smaller
value of $L_z$ than the lower, with the consequence that on it $\theta_1'$
librates around $\pi/2$ rather than zero.}\label{fig:full_int_orb}
\end{figure}

\begin{figure}
\begin{center}
\includegraphics[width=.9\hsize]{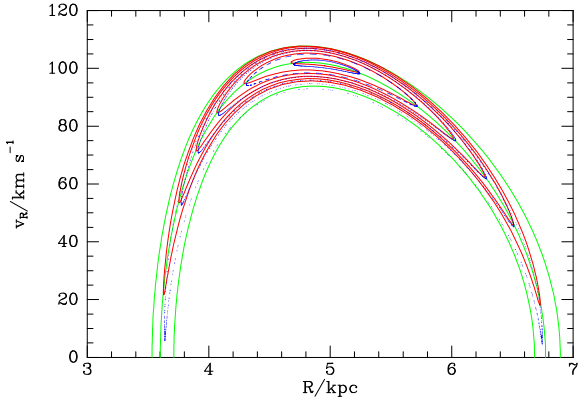}
\end{center}
 \caption{Red curves: sections through trapped tori with a low value of $L_z$
constructed by perturbation theory. Blue points: consequents of numerically
integrated orbits started from a point on each perturbatively constructed
torus. Green curves: untrapped tori fitted by \TM\ either side of the
trapping region and the resonant torus that is constructed by interpolating
between these fitted tori.}\label{fig:full_interp0}
\end{figure}

\begin{figure}
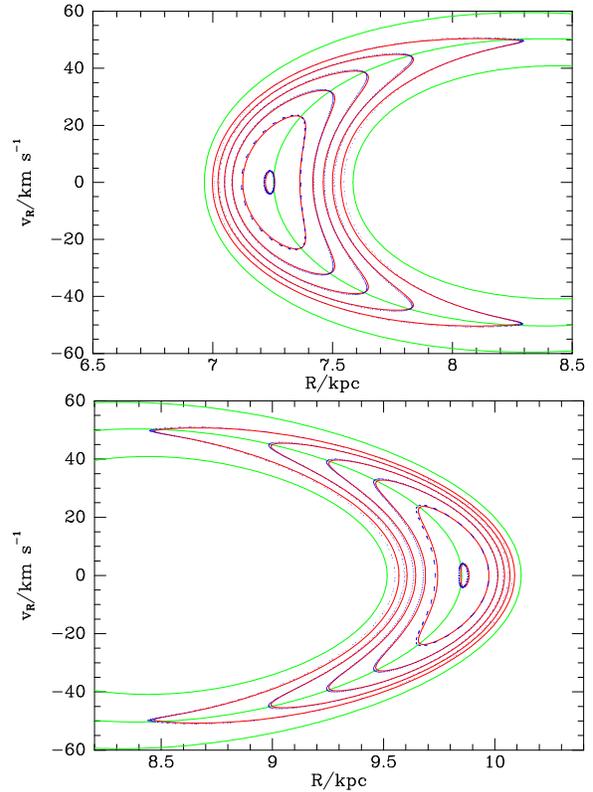

\begin{center}
\includegraphics[width=.9\hsize]{plots/full_interp1.ps}
\includegraphics[width=.9\hsize]{plots/full_interp1r.ps}
\end{center}
 \caption{As \figref{fig:full_interp0} but for orbits with larger values of
$L_z$, so trapping occurs around $\theta_1'=0$ (top) or $\theta_1'=\pi$
(bottom).}\label{fig:full_interp1}
\end{figure}

Once the values on the resonant torus of $h_2$ and $h_4$ and their first two
derivatives are known, a trapped torus can be quickly constructed for any
allowed value of $I_1$. \figref{fig:full_int_orb} shows two trapped orbits
reconstructed from perturbation theory in red and from integration of the
full equations of motion in dashed blue lines. The agreement is excellent.
The red curves in \figref{fig:full_interp0} show sections through tori
trapped around $\theta_1'=\pi/2$ constructed by perturbation theory for six
equally spaced values of $I_1$. The blue curves in \figref{fig:full_interp0}
show consequents of numerically integrated orbits with initial conditions
drawn from each perturbatively constructed torus. The agreement is good.  In
\figref{fig:full_interp0} we show in green the untrapped tori fitted on
either side of the trapping region, and the resonant torus constructed by
interpolating between these fitted tori.

\figref{fig:full_interp1} shows surfaces of section for orbits with larger
values of $L_z$ that are trapped around $\theta'_1=0$ (upper panel) and
$\theta_1'=\pi$ (lower panel). The agreement between the red curves from
perturbation theory and the blue consequents from full orbit integration is
almost perfect.  When $\theta'_1$ librates around $\pi$ (lower panel) the
motion in the $Rz$ plane is just the time reverse of motion on the orbit with
the same value of $I_1$ that is trapped around $\theta_1'=0$ (upper panel)
with the consequence that the orbits look identical in a plot such as
\figref{fig:full_int_orb}. 

To construct a red curve in \figref{fig:full_interp0} or
\ref{fig:full_interp1}, one chooses a value for $I_1$ and for that value \TM\
determines the permitted range in $\theta_1'$ (eq.~\ref{eq:discr}). Then for
values of $\theta'_1$ in the permitted range \TM\ computes $J'_1$ from
equation \eqref{eq:Hquad}. Next adopting the value taken by $J'_2$ on the
resonant torus \TM\ computes $(J_r,J_z)$ from equations \eqref{eq:defsJp}.
\TM\ generates the surface of section by varying $\theta_z$ with
$\theta_r-\theta_z$ fixed at the chosen value of $\theta'_1$ until $z=0$.

The torus corresponding to the outermost red curve in
\figref{fig:full_interp0} has action of libration $\cJ=7.37\kpc\kms$. Twice
this action agrees quite well with the width ($16.6\kpc\kms$) of the ZoMA
determined in Section~\ref{sec:resTM} from orbit integrations and untrapped
tori from \TM.  This agreement both confirms the accuracy of the numerical
work and assures us that negligible phase-space volume is taken up with
orbits that we have not considered.

\begin{figure}
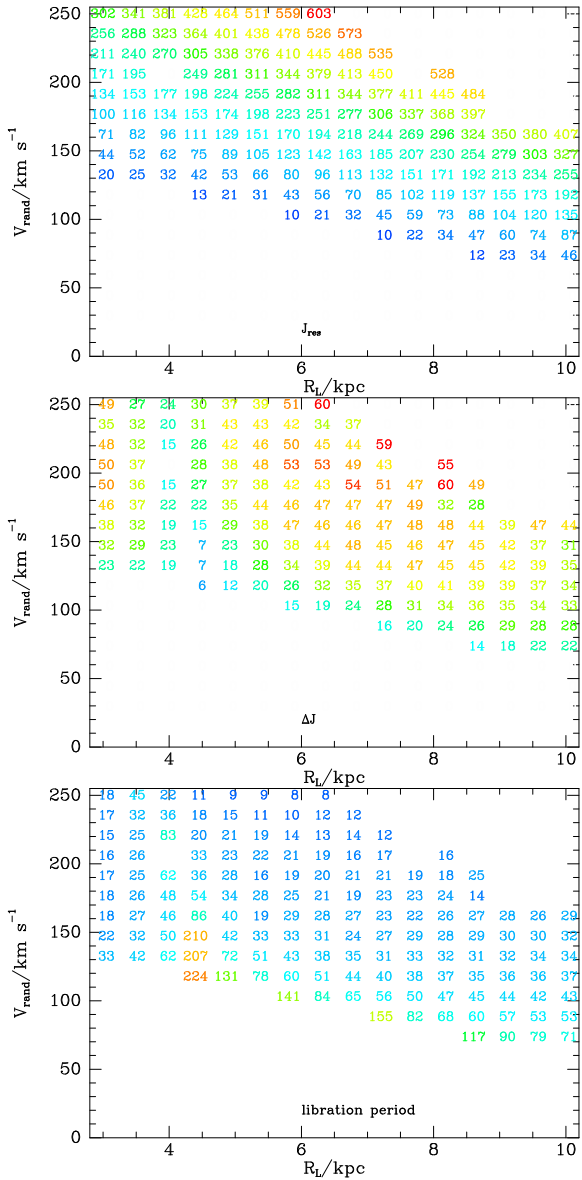

\begin{center}
\includegraphics[width=.9\hsize]{plots/pljres1.ps}
\includegraphics[width=.9\hsize]{plots/pljres0.ps}
\includegraphics[width=.9\hsize]{plots/nv0.ps}
\end{center}
\caption{A representation of the $(E,L_z)$ plane showing the extent of
resonantly trapped orbits. The horizontal axis quantifies $L_z$ in terms of
the radius $\RL\equiv L_z/250\kms$. The vertical axis quantifies
the amount by which an orbit's energy exceeds that of the circular orbit in
terms of the corresponding speed in the $Rz$-plane. The numbers in the top
panel give, in units of $\kpc\kms$, the radial action $J_r(\hbox{res})$ of
the resonant torus with the given $(L_z,E)$. The figures in the middle panel
give the width of the band of $J_r$ centred on $J_r(\hbox{res})$ within which
orbits are trapped. The numbers in the bottom panel give the libration period
in units of $100\Myr$ of the trapped orbit with $I_1=(I_{1\,\rm max}+I_{1\,\rm min})/2$.
The numbers are colour-coded to make large-scale trends
apparent.}\label{fig:pljres}
\end{figure}

\subsection{Extent of trapped orbits}

What is the extent of resonant trapping in the Galaxy's phase space? How
large are the errors that will arise if we neglect this phenomenon?

\figref{fig:pljres} displays the results of a survey of action space for the
extent of resonant trapping. Horizontally we plot $L_z$ through the radius
$\RL\equiv L_c/250\kms$, which is approximately the radius of the
corresponding circular orbit. Vertically we plot the quantity $v_{\rm rand}$
defined by
\[
\fracj12v_{\rm rand}^2\equiv E-E_{\rm c},
\]
 where $E_{\rm c}$ is the energy of the circular orbit with the given angular
momentum. So $v_{\rm rand}$ is the speed at which the orbit passes through
its guiding centre. The numbers written in the top panel give the value of
$J_r$ (in units of $\!\kpc\kms$) of the resonant torus with the given
$(\RL,v_{\rm rand})$. The numbers in the middle panel give the corresponding
width $\Delta J_r$ of the trapping region region. The numbers in the bottom
panel give, in units of $100\Myr$, the libration period in the middle of the
trapping region, i.e., for the orbit with $I_1=(I_{1\,\rm max}+I_{1\,\rm
min})/2$. The blank region in the
lower part of each panel indicates that for small values of $v_{\rm rand}$
(orbits that are neither highly eccentric nor steeply inclined to the plane)
$\Omega_r<\Omega_z$ so resonant trapping cannot occur. For a given value of
$\RL$, there is a critical value of $v_{\rm rand}$ at which trapping occurs,
for a few orbits (small $\Delta J_r$) and at small values of $J_r$ (top
panel). That is, trapping commences with orbits close to a shell orbit. The
critical value of $v_{\rm rand}$ for trapping decreases roughly linearly with
increasing $\RL$. As $v_{\rm rand}$ increases at fixed $\RL$, the values of
both $J_r$ and $J_z$ on the resonant orbit rise (top panel), and the
trapping region widens (middle panel). The width, however, quickly becomes
small compared to $J_r$, so most orbits remain untrapped. 

The libration period (bottom panel) is less than $1\Gyr$ for only a few
orbits near the top of the trapping region, and it increases to over $20\Gyr$
along the bottom of the trapping region near where the sign of $h_{2,2}$
changes.

\begin{figure}
\begin{center}
\includegraphics[width=.9\hsize]{plots/chaos.ps}
\end{center}
\caption{A chaotic orbit associated with $(\RL=8\kpc,v_{\rm rand}=210\kms)$
in \figref{fig:pljres}.}\label{fig:chaos}
\end{figure}

In each panel of \figref{fig:pljres} the block of numbers for trapped orbits
has an upper edge that slopes down from left to right. This boundary is
associated with the onset of chaos, as \figref{fig:chaos} illustrates by
showing an orbit associated with the upper edge of the column of numbers at
$\RL=8\kpc$ in \figref{fig:pljres}. This numerically integrated orbit clearly
consists of sections in which the star is resonantly trapped around
$\phi_1'=\pm\pi/2$ joined by sections of untrapped motion. In the portion of
phase space occupied by these highly eccentric and mildly chaotic orbits,
\TM\ returns values of the perturbing term in the Hamiltonian $h_2$ that
fluctuate erratically in sign: the sign of $h_2$ must depend sensitively on
the untrapped tori used to generate the interpolated tori on which the
Hamiltonian is Fourier analysed. As a consequence of these sign changes, the
survey program that produced \figref{fig:pljres} reports that trapping does
not occur.

In the lower panel of \figref{fig:pljres} the column of values of $\Delta
J_r$ at $\RL=4\kpc$ shows anomalously small values. This phenomenon reflects
the change in the phase $\psi_2$ from $\pm\pi$ at small $\RL$ to $0$ at
larger $\RL$, which requires $h_2$ to vanish at a critical value of $\RL$.

\subsection{Effect of ignoring resonant trapping}

\figref{fig:pljres} shows that resonant trapping will not affect thin-disc
stars because it requires random velocities $\ga80\kms$ for stars with
guiding centres near the Sun, and even larger random velocities for stars
with smaller guiding centres. Trapping is principally an issue for halo
stars, but also for some thick-disc stars.

Figs.~\ref{fig:full_int_orb} to \ref{fig:full_interp1} show excellent
agreement between numerically computed orbits and the results of perturbation
theory in which stars move slowly between untrapped tori. Hence the tori
constructed by \TM\ provide appropriate velocities at all points in phase
space, and for the Galaxy modeller, the only issue is how to weight those
velocities when predicting observables. Ideally a \df\ $f(\vx,\vv)$ would
separately specify the weight of each trapped orbit and each untrapped orbit
and use these values to weight separately velocities occurring on untrapped
and trapped orbits. In practice we rely on a \df\ $f(\vJ)$ that just specifies the
weights of untrapped orbits and uses these to weight velocities regardless of
whether they occur on trapped or untrapped orbits. 

We could, in principle, grossly over- or under-populate trapped orbits by
this procedure, because trapped orbits could conceivably be bereft of stars
or be exceptionally heavily populated by stars. From our current  perspective
neither hypothesis is more likely than the other: we are restricting
ourselves to consideration of dynamically consistent models of the Galaxy as
it currently is, and the question of how it arrived in this configuration,
upon which the population of trapped orbits depends, lies beyond our present
scope. An intriguing conjecture that suggests that trapped orbits may be
anomalously populated is the ``levitation'' model of \cite{SridharTouma}.

Here we simply investigate the well-defined hypothesis that the value of $f$
on a trapped orbit is the angle (time) average $\overline{f}$ along that
orbit of the value of the \df\ $f(\vJ)$ of the untrapped orbits. This
hypothesis is a plausible one, and would be essentially true if the intensity
of scattering of stars by fluctuations in the Galaxy's gravitational field
were high enough for stars to have a good change of being scattered into and
out trapped orbits within a Hubble time. We will show that if this hypothesis
is correct, only small errors will be made when resonant trapping is simply
ignored.

\begin{figure}
\begin{center}
\includegraphics[width=.9\hsize]{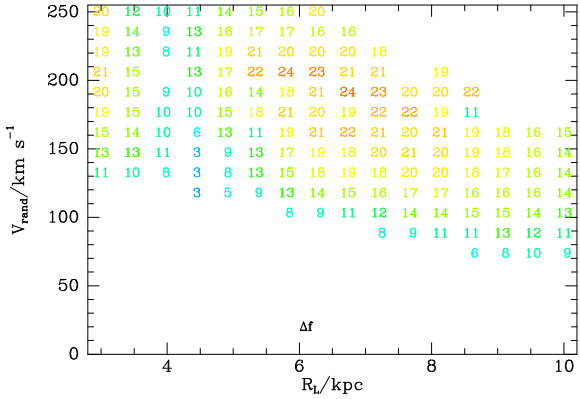}
\end{center}
\caption{The upper limits (in percent) on the rms variation around resonantly trapped
orbits of a realistic disc \df. The \df\ is that fitted by Piffl et al.\
(2014) to data from RAVE and SDSS.}\label{fig:varf}
\end{figure}

Under our hypothesis that $\overline{f}$ is the correct \df\ for trapped
orbits, an indicator of the errors introduced by ignoring trapping is the rms
value of $f(\vJ)-\overline{f}$ along a resonantly trapped orbit. This rms
increases from very small values for trapped orbits with $I_1=I_{1\,\rm
max}$, to a maximum value on orbits with $I_1=I_{1\,\rm min}$.
\figref{fig:varf} shows the rms variation around trapped orbits with
$I_1=I_{1\,\rm min}$ of the \df\ for the Galactic disc that \cite{Piea14}
fitted to terminal velocities, RAVE kinematics and SDSS star counts. Most of
the (maximal) rms variations plotted in \figref{fig:varf} are smaller than 20
per cent, although a handful range up to 24 per cent. Consequently, if we
ignore resonant trapping, we will make errors in the density of stars in
velocity space that can be as high as 24 per cent at certain velocities. But
in general our error will be much smaller because: (i) most trapped orbits
have $I_1>I_{1\,\rm min}$; (ii) even when $(\RL,v_{\rm rand})$ are such that
entrapment is possible, the great majority of orbits are not resonantly
trapped because the width $\Delta J_r$ of the region of entrapment is small
compared to $J_r$ on the resonant orbit; (iii) at small values of $v_{\rm
rand}$ no orbits are trapped.

\subsection{Computing observables with entrapment included}

\begin{figure}
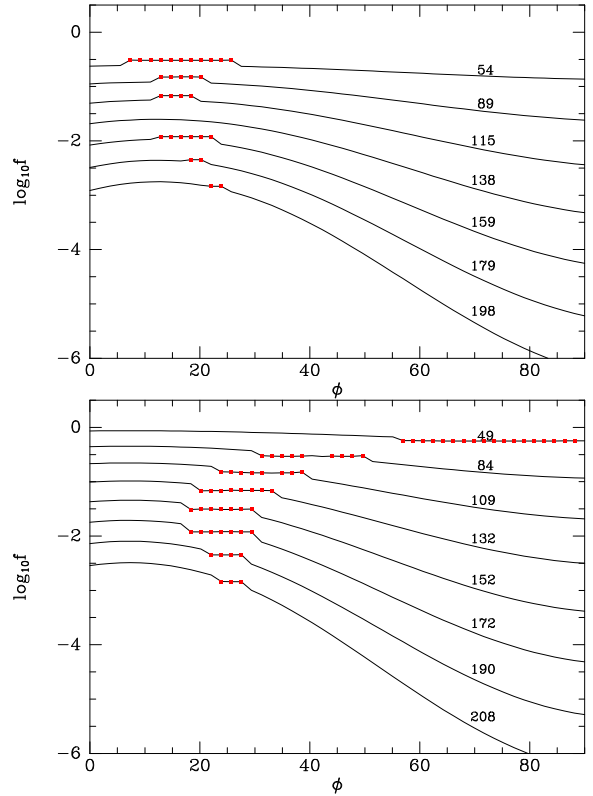

\begin{center}
\includegraphics[width=.9\hsize]{plots/dfbar.ps}
\includegraphics[width=.9\hsize]{plots/dfbar2.ps}
\end{center}
\caption{Density of stars in velocity space at $(R,z)=(8,2)\kpc$ (upper
panel) and $(R,z)=(8,1.2)\kpc$ (lower panel). Each curve
shows the value of a realistic disc \df\ on orbits of a given energy and
angular momentum as a function of the angle between the velocity vector and
the $R$ axis as the orbit passes through the given point in the $Rz$ plane.
All orbits have angular
momentum
$L_z=4.9\kpc\times250\kms$. Each curve is labelled with the
orbit's speed (in $\!\kms$) at the given location. The red dots are associated with
resonantly trapped orbits, and the \df\ on these orbits is set equal to the
time-averaged value of the \df\ for non-resonant orbits.}\label{fig:dfbar}
\end{figure}

Consider now how to compute the star density in velocity space at $(R,z)$
when one acknowledges trapping and weights trapped orbits with
$\overline{f}$.  From $(R,z,\vv)$ we easily compute $\RL(J_\phi)$ and $v_{\rm
rand}$.  If $v_{\rm rand}$ does not lie in the band of entrapment in
\figref{fig:pljres}, entrapment is not an issue. If it does lie in this band,
we find in tables constructed during the survey of action space that yielded
\figref{fig:pljres} the corresponding values of the actions $\vJ_0$ of the
resonant torus, and the perturbation parameters $G$ and $h_i$.

From the St\"ackel Fudge \citep{JJB12:Stackel,SaJJB16} and $(R,z,\vv)$ we
obtain $(\vtheta,\vJ)$. We correct $\vJ$ to ensure that $J_2'\equiv J_r+J_z$
agrees with the value assigned by \TM\ to the given angular momentum and
energy. We determine whether $J_r$ lies within the range of trapped actions
implied by $\vJ_0$, $G$ and $h_i$. If it does not, trapping is not an issue.
When $J_r$ does lie in the trapping range, we compute $\Delta=J_r-J_{0r}$ and
$\theta_1'=\theta_r-\theta_z$. With these values we can compute $I_1$ from
equation \eqref{eq:Hquad}. If $I_1$ lies between the minimum and maximum
values of $I_1$ for trapped orbits, we find $\overline{f}(\RL,v_{\rm
rand},I_1)$ by interpolating in tables of the time-averaged mean of $f(\vJ)$
on trapped orbits. If $I_1<I_{1\,\rm min}$ or $I_1>I_{1\,\rm max}$, trapping
is not an issue and we simply evaluate $f(\vJ)$. 

\figref{fig:dfbar} shows the resulting values of the \df\ in the velocity
spaces at $(R,z)=(8,2)\kpc$ and $(8.1.2)\kpc$ in the upper and lower panels,
respectively. All orbits have the same angular momentum, that associated
with $R_L=4.9\kpc$, and each curve shows the value of the \df\ on orbits of a
given energy, namely that associated with the speed at the given location marked above
each curve. Orbits marked with red points are resonantly
trapped, and one sees  that their phase-space densities differ from
those of untrapped orbits. The effect is, however, slight. Moreover, at
$z\ga1.8\kpc$ trapping is confined to low angular momenta.

Although we do not use energy as an argument of the \df, much of the
dependence of $f$ on $\vJ$ is through the Hamiltonian $H(\vJ)$. Librating
stars move on surfaces of constant $\overline{H}$, so  they move on surfaces
in action space characterised by only weak variations in $f(\vJ)$. In
particular, if the velocity ellipsoids have nearly round projections onto
the $v_Rv_z$ plane, librating stars will not stray far from their original
surface of constant $f(\vJ)$.

In this example the error one makes by ignoring trapping is small. This is a
significant conclusion because it validates previous work modelling the
Galaxy with \df s of the form $f(\vJ)$ by showing that  models exist in which
trapping is correctly deal with, that would yield very similar observables.
However, it does not exclude the existence of other models in which trapping
is correctly handled that would have significantly different observables
despite having the same \df\ for all untrapped orbits. These other models
would weight trapped orbits with a \df\ very unlike $\overline{f}$. We leave
for a future study deciding whether such models really exist, and if so are relevant
for our Galaxy.

\section{Conclusions}\label{sec:conclude}

Global angle-action coordinates are hugely convenient tools for galaxy
modelling, and real progress in understanding our Galaxy has recently been
achieved through their use. We anticipate that angle-action coordinates will
also prove valuable for modeling external galaxies. Unfortunately they are an
idealisation since in real gravitational potentials some orbits become
resonantly trapped. Trapped orbits do have action integrals, but some of
these do not fit into the framework set by the untrapped orbits.

In any axisymmetric model of a disc galaxy the resonance $\Omega_r=\Omega_z$
must always be considered.  We have elucidated the behaviour of the code \TM\
in regions of action space affected by resonant trapping. With large values
of the code's tolerance parameter $\tolJ$, \TM\ can be used to foliate phase
space with tori regardless of resonant trapping. However, when small values
of $\tolJ$ are used, the structure of the tori \TM\ constructs changes
abruptly near a resonant torus, with the result that adjacent tori are liable
to cross.  Crossing of tori is to be avoided. We do so by using \TM\ to
compute a sparse grid of (untrapped) tori and then using linear interpolation
to foliate phase space with tori. These tori define an integrable Hamiltonian
that is close to the true Hamiltonian, so first-order perturbation theory can
be used to compute trapped orbits with good precision. Comparison of the
\sos\ in \figref{fig:full_interp0} with Fig.~3.38 in \cite{GDII}, which was
constructed by standard Hamiltonian perturbation theory, demonstrates the
dramatic increase in precision that \TM\ enables.

We have extended the code \TM\ to manipulate orbits trapped by the resonance
$\Omega_r=\Omega_z$ as conveniently as one can manipulate untrapped orbits.
We have used the extended code to investigate the extent of trapping by this
resonance in a realistic model of our Galaxy's potential. The extent is not
major but possibly large enough to be of astronomical significance, if only
because it permits net streaming of stars within the $(R,z)$ plane. Trapping
is confined to stars on significantly non-circular orbits, so it has
negligible impact on the thin disc: to become trapped a star needs to move
faster than $\sim80\kms$ with respect its guiding centre.

Trapped orbits require a \df\ that is distinct from that used to
populate untrapped orbits. However, a natural choice of  \df\ is the time
average along each trapped orbit of the \df\ for untrapped orbits. We have
investigated the density of stars in phase space at locations dominated by
thick-disc stars when this ansatz is used to populate trapped orbits using
the \df\ for our Galaxy that \cite{Piea14} fitted to a large body of data.
We find that  trapping then has only a small impact on the structure of
velocity space.

In the near future, precise kinematic data from Gaia and
spectra from ground-based surveys should enable us to constrain the Galaxy's
potential so tightly that we will be sure which parts of local velocity space
correspond to trapped orbits. Then it will be important to seek evidence of
different phase-space densities on orbits trapped at the two relevant phases,
and abrupt changes in the density of stars across the edges of regions of
entrapment. The precise location in velocity space of these changes would
constitute a useful constraint on the potential.

At the present time it might be argued that there are more urgent tasks for
Galaxy modellers than devising entertaining \df s for resonantly trapped
orbits. Nevertheless, one cannot help being curious about models in which
all orbits trapped by the resonance $\Omega_r=\Omega_z$ circulate in the
$Rz$ plane in one sense. Here we have laid out the tools required to build
such a model. Another entertaining exercise would be to apply these tools to
the cubic Galaxy potential introduced by \cite{HenonHeiles} and using them study
the onset of chaos in this potential as the energy approaches the escape
energy. 

Perhaps a more useful direction for further work is to extend torus mapping
to rotating barred potentials and to use techniques similar to those deployed
here to model the series of resonances that cause phase space to break up as
one approaches the corotation resonance.

\section*{Acknowledgements}

I am grateful to the referee and members of the Oxford Galaxy dynamics group
for helpful comments on drafts.  This work was supported by the UK Science
and Technology Facilities Council (STFC) through grant ST/K00106X/1 and by
the European Research Council under the European Union's Seventh Framework
Programme (FP7/2007-2013)/ERC grant agreement no.~321067.

\bibliographystyle{mn2e} \bibliography{/u/tex/papers/mcmillan/torus/new_refs}

\begin{thebibliography}{}

\bibitem[\protect\citeauthoryear{{Binney}}{{Binney}}{2010}]{JJB10}
{Binney} J.,  2010, \mnras, 401, 2318

\bibitem[\protect\citeauthoryear{{Binney}}{{Binney}}{2012}]{JJB12:Stackel}
{Binney} J.,  2012, \mnras, 426, 1324

\bibitem[\protect\citeauthoryear{{Binney} \& {McMillan}}{{Binney} \&
  {McMillan}}{2011}]{JJBPJM11:dyn}
{Binney} J.,  {McMillan} P.,  2011, \mnras, 413, 1889

\bibitem[\protect\citeauthoryear{{Binney} \& {McMillan}}{{Binney} \&
  {McMillan}}{2016}]{JJBPJM16}
{Binney} J.,  {McMillan} P.~J.,  2016, \mnras, 456, 1982

\bibitem[\protect\citeauthoryear{{Binney} \& {Piffl}}{{Binney} \&
  {Piffl}}{2015}]{BinneyPiffl2015}
{Binney} J.,  {Piffl} T.,  2015, \mnras, p.~xx

\bibitem[\protect\citeauthoryear{{Binney} \& {Spergel}}{{Binney} \&
  {Spergel}}{1984}]{BinneySII}
{Binney} J.,  {Spergel} D.,  1984, \mnras, 206, 159

\bibitem[\protect\citeauthoryear{{Binney} \& {Tremaine}}{{Binney} \&
  {Tremaine}}{2008}]{GDII}
{Binney} J.,  {Tremaine} S.,  2008, {Galactic Dynamics: Second Edition}.
Princeton University Press

\bibitem[\protect\citeauthoryear{{Chirikov}}{{Chirikov}}{1979}]{Chirikov1979}
{Chirikov} B.~V.,  1979, Phys. Rep., 52, 263

\bibitem[\protect\citeauthoryear{{Dehnen} \& {Binney}}{{Dehnen} \&
  {Binney}}{1998}]{WDJJB98:Mass}
{Dehnen} W.,  {Binney} J.,  1998, \mnras, 294, 429

\bibitem[\protect\citeauthoryear{{Henon} \& {Heiles}}{{Henon} \&
  {Heiles}}{1964}]{HenonHeiles}
{Henon} M.,  {Heiles} C.,  1964, \aj, 69, 73

\bibitem[\protect\citeauthoryear{{Kaasalainen}}{{Kaasalainen}}{1994}]{Kaasalainen_res}
{Kaasalainen} M.,  1994, \mnras, 268, 1041

\bibitem[\protect\citeauthoryear{{Kaasalainen} \& {Binney}}{{Kaasalainen} \&
  {Binney}}{1994}]{KaJJB94:PhysRev}
{Kaasalainen} M.,  {Binney} J.,  1994, Physical Review Letters, 73, 2377

\bibitem[\protect\citeauthoryear{{McMillan}}{{McMillan}}{2011}]{PJM11:mass}
{McMillan} P.~J.,  2011, \mnras, 414, 2446

\bibitem[\protect\citeauthoryear{{Piffl}, {Binney}, {McMillan}, {Bienaym{\'e}},
  {Bland-Hawthorn}, {Freeman}, {Gibson}, {Gilmore}, {Grebel}, {Helmi},
  {Kordopatis}, {Navarro}, {Parker}, {Reid}, {Seabroke}, {Siebert}, {Steinmetz}
  \& {Wyse}}{{Piffl} et~al.}{2014}]{Piea14}
{Piffl} T.,  {Binney} J.,  {McMillan} P.~J.,  {Bienaym{\'e}} O.,
  {Bland-Hawthorn} J.,  {Freeman} K.,  {Gibson} B.,  {Gilmore} G.,  {Grebel}
  E.~K.,  {Helmi} A.,  {Kordopatis} G.,  {Navarro} J.~F.,  {Parker} Q.,  {Reid}
  W.~A.,  {Seabroke} G.,  {Siebert} A.,  {Steinmetz} M.,    {Wyse} R.~F.~G.,
  2014, \mnras, 445, 3133

\bibitem[\protect\citeauthoryear{{Press}, {Flannery} \& {Teukolsky}}{{Press}
  et~al.}{1986}]{NumRec}
{Press} W.~H.,  {Flannery} B.~P.,    {Teukolsky} S.~A.,  1986, {Numerical
  recipes. The art of scientific computing}.
Cambridge: University Press, 1986

\bibitem[\protect\citeauthoryear{{Sanders} \& {Binney}}{{Sanders} \&
  {Binney}}{2013}]{SaJJB13b}
{Sanders} J.~L.,  {Binney} J.,  2013, \mnras, 433, 1826

\bibitem[\protect\citeauthoryear{{Sanders} \& {Binney}}{{Sanders} \&
  {Binney}}{2015}]{SaJJB15:EDF}
{Sanders} J.~L.,  {Binney} J.,  2015, ArXiv e-prints

\bibitem[\protect\citeauthoryear{{Sanders} \& {Binney}}{{Sanders} \&
  {Binney}}{2016}]{SaJJB16}
{Sanders} J.~L.,  {Binney} J.,  2016, \mnras, 457, 2107

\bibitem[\protect\citeauthoryear{{Sch{\"o}nrich} \& {McMillan}}{{Sch{\"o}nrich}
  \& {McMillan}}{2016}]{SchoenPJM}
{Sch{\"o}nrich} R.,  {McMillan} P.,  2016, ArXiv e-prints

\bibitem[\protect\citeauthoryear{{Sridhar} \& {Touma}}{{Sridhar} \&
  {Touma}}{1996}]{SridharTouma}
{Sridhar} S.,  {Touma} J.,  1996, \mnras, 279, 1263

\end{thebibliography}

\appendix
\section{Are resonant coefficients well defined?}\label{app:A}

\TM\ uses the Levenberg-Marquardt
algorithm to minimise $\ex{H_1^2}_\vtheta$. This involves computing an
approximation to the matrix
\[
D_{ij}\equiv{\p^2\!\ex{H_1^2}\over\p a_i\p a_j}
\]
 of second derivatives of $\ex{H_1^2}$ with respect to the parameters $a_i$,
and then reasoning that near the minimum of $\ex{H_1^2}$ we
have
\[
\ex{H_1^2}\simeq\hbox{const}+\fracj12\sum_{ij}D_{ij}(a_i-a_{{\rm
min},i})(a_j-a_{{\rm min},j})
\]
 where $\va_{\rm min}$ is the set of parameter values that minimises
 $\ex{H_1^2}$. Hence in the vicinity of this minimum,
\[
{\p\ex{H_1^2}\over\p a_i}\simeq \sum_{j}D_{ij}(a_j-a_{{\rm min},j}).
\]
Finally a modified Newton-Raphson algorithm is used to solve these equations
for $\va-\va_{\rm min}$ given the computed value of the gradient on the left.
One might worry that the divergence of the resonant $S_\vn$ as one approaches
the centre of the  ZoMA arises because $\vD$ is tending to  a
singular matrix, and that  the parameters $a_i$ that correspond to resonant $S_\vn$
lie in the kernel of $\vD$. We now show that this is not the case: all $a_i$
remain well defined.

\begin{figure}
\centerline{\includegraphics[width=.9\hsize]{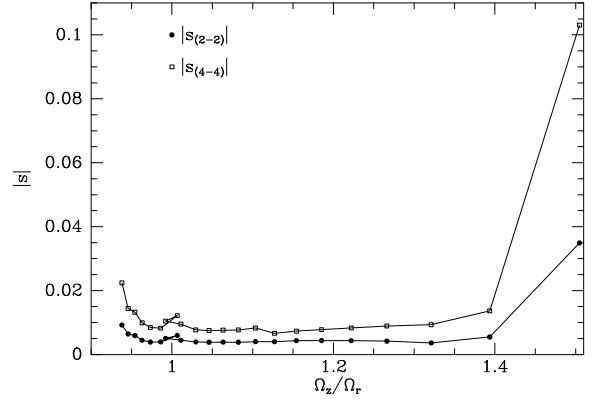}}
\caption{The sensitivities $|\vs|$ (eqn \ref{eq:sensitive}) of the
coefficients  $S_{(2-2)}$ (filled points) and $S_{(4-4)}$ (open
points) on tori constructed by \TM\ for the surface of section
\figref{fig:four} plotted against frequency ratio.}\label{fig:sensitive}
\end{figure}

To investigate this hypothesis we compute the singular-value decomposition of
$\vD$. That is, we compute orthogonal matrices $\vU$ and $\vV$ and a diagonal
matrix $\vW$ such that \citep{NumRec}
\[
\vD=\vU\cdot\vW\cdot\vV^T.
\]
 If $\vD$ becomes singular, some of the diagonal elements of $\vW$ tend to
zero. If $W_{ii}\to0$, the $i$th column vector of $\vV$ makes a vanishing
contribution to $\vD$, with the consequence that it enters the kernel of
$\vD$. So if the resonant $S_\vn$ enter
the kernel of $\vD$, they will be projected onto column vectors of $\vV$ that
correspond to vanishing $W_{ii}$. In \figref{fig:sensitive} we plot the
moduli of the vectors
\[\label{eq:sensitive}
s_i\equiv\sum_jV_{ij}W_{jj}
\]
for the values of $i$ that correspond to $S_{(2,-2)}$ and $S_{(4,-4)}$. We see
that these moduli are small not just at the resonance, but quite far from it.
Thus the growth of the resonant $S_\vn$ does not arise because $\vD$
becomes singular with the resonant coefficients in its kernel.

\begin{figure}
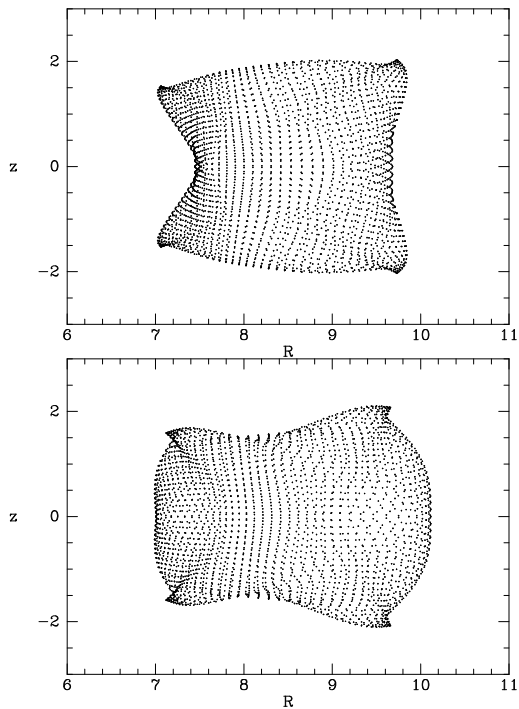

\begin{center}
\includegraphics[width=.8\hsize]{plots/net0.ps}
\includegraphics[width=.8\hsize]{plots/net1.ps}
\end{center}
\caption{Grids of $80\times80$ points uniformly distributed in toy angles on tori fitted
by \TM\ on either side of the resonance. The upper panel is for the smaller
value of $J_r$.}\label{fig:net}
\end{figure}

Given that $\vD$ is never a singular matrix, how should we understand the
discontinuous change in the resonant coefficients in the ZoMA? 

If we were to
plot contours of constant $\ex{H_1^2}$ in the space spanned by the parameters
$\va$, we would sometimes find multiple local minima. \TM's job is to locate
the deepest minimum, and our experience is that it does this job remarkably
reliably. When the prescribed actions do not lie in the ZoMA, it finds the
minimum that yields very small $\ex{H_1^2}$, so the associated angle-action
coordinates are very close to those of the true Hamiltonian. As the specified
actions move into the ZoMA, the chosen minimum has a
distinctly non-zero value of $\ex{H_1^2}$. As we move deeper into the ZoMA,
this value becomes ever larger. Meanwhile the value of $\ex{H_1^2}$
associated with some other local minimum is falling, and eventually drops
below that of the minimum that \TM\ has been finding.  Around this time of
crossover in the values of $\ex{H_1^2}$ at the minima, \TM\
starts finding the minimum that has the falling value, and as the actions
emerge from the ZoMA, the value of $\ex{H_1^2}$ at this minimum approaches
zero.

\figref{fig:net} shows for tori that lie either side of the crossover, grids
of  points in real space at which \TM\ has evaluated the true Hamiltonian. On
account of the large magnitude of the resonant coefficients $S_\vn$ and
opposite signs on each side of the transition, the two fitted tori have
markedly different spatial extents. Hence the equations that lead to the two
minima are obtained by evaluating the Hamiltonian at significantly different
phase-space points, and  the suddenness of
the transition  in \figref{fig:Sn} is a consequence.

\subsection{Other discontinuities}

\begin{figure}
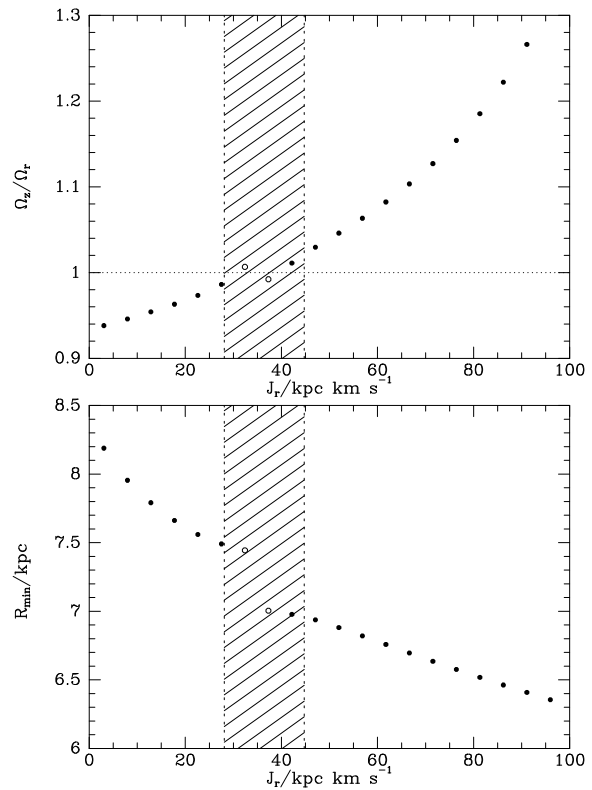

\begin{center}
\includegraphics[width=.9\hsize]{plots/omegaJ.ps}
\includegraphics[width=.9\hsize]{plots/RminJ.ps}
\end{center}
\caption{Top: The frequency ratio $\Omega_z/\Omega_r$ as a function of $J_r$:
filled dots are for  tori shown in the surface of section \figref{fig:four};
open dots are for two additional tori that \TM\ 
returned with error flag $-2$. Below: The
minimum value of $R$ on these tori. In each panel the ZoMA is
shaded.}\label{fig:res}
\end{figure}

In the upper panel of \figref{fig:res} we plot the ratio of frequencies
$\Omega_z/\Omega_r$ for tori returned by \TM\ at the energy of the \sos\ of
\figref{fig:four}. The ZoMA is cross-hatched. Two of the dots in the ZoMA are
for tori that \TM\ returned with non-zero error flag, implying that the
target value of $\ex{H_1^2}$ was not attained. The cross-sections of all
the tori associated with all other dots are plotted in \figref{fig:four}. 

The lower panel of \figref{fig:res} shows the minimum radii $R_{\rm min}$ at which the
returned tori intersect the surface $z=0,v_R=0$. Within the ZoMA there is an
abrupt change in $R_{\rm min}$ associated with the change in sign of the
resonant $S_\vn$. The trapped orbit that forms the boundary of the
island of resonant trapped orbits in \figref{fig:three} moves between tori
that vary in actions between $28.13\kpc\kms$ and $44.74\kpc\kms$. 

\end{document}